# Single Harmonic-based Narrowband Magnetic Particle Imaging

*Klaas-Julian Janssen, Meinhard Schilling, Frank Ludwig and Jing Zhong\**

Institute for Electrical Measurement Science and Fundamental Electrical Engineering and Laboratory for Emerging Nanometrology (LENA), TU Braunschweig, Hans-Sommer-Str. 66, Braunschweig D-38106, Germany.

*Corresponding author: j.zhong@tu-braunschweig.de

*Abstract*—**Visualization of the in vivo spatial distribution of superparamagnetic iron oxide nanoparticles (SPIONs) is crucial to biomedicine. Magnetic particle imaging (MPI) is one of the most promising approaches for direct measurements of the SPION distribution. In this paper, we systematically investigate a single-harmonic-based narrowband MPI approach. Herein, only the 3$^{rd}$ harmonic at 15 kHz of the SPION signal induced in an excitation magnetic field of 5 kHz is measured via a narrowband detection system for imaging during scanning a field-free-point in a field of view. Experiments on spot and line phantoms are performed to evaluate the spatial distribution by the assessment of the full width at half maximum and modulation transfer function at different excitation magnetic fields from 4 to 10 mT. Experimental results demonstrate that reconstructed images have a spatial resolution of 1.6 and 1.5 mm for a gradient field of 2.2 T/m and 4.4 T/m in x- and z-direction, respectively, at an excitation magnetic field of 4 mT. In terms of line gap, two lines with a gap of 0.5 mm are resolved. With increasing the excitation magnetic field to 10 mT, the spatial resolution gets worse to 2.4 and 2.0 mm in x- and z-direction, respectively. Moreover, the custom-built MPI scanner allows a limit of detection of 53 µg (Fe)/mL (500 ng Fe weight) using perimag SPIONs. In addition, the excellent performance is demonstrated by imaging experiments on an "emg" logo phantom. We believe that the proposed narrowband MPI approach is a promising approach for SPION imaging.**

## I. INTRODUCTION

SUPERPARAMAGNETIC iron oxide nanoparticles (SPIONs) have shown great promise as an emerging platform in disease diagnostics and therapy. SPIONs functionalized with antibodies bind to their specific antigens, allowing the tracking and imaging of the antigens for disease diagnostics [1]–[4]. In targeted drug delivery, SPIONs together with drugs capsulated into polymers are used as magnetic carriers for targeting specific tissue with external gradient magnetic fields to improve treatment efficiency [5]–[7]. In addition, the SPIONs are also used as heaters when applying radio-frequency ac magnetic fields to release the drugs from the polymers [8]–[11]. In cell-based cancer therapy, e.g. adoptive cellular therapy, the labelling of SPIONs onto the surface of specific cells allows in vivo tracking [12]–[16], which is of great importance to understand proliferation of the adoptive cells. For all these applications, a novel imaging modality for the SPION imaging provides new insights into the development of these techniques for disease diagnostics and therapy, including the measurements of the specific antigens for diagnostics, active targeting of the drugs to specific tissues for therapy and understanding of the adoptive cell transplantation [17], [18]. Therefore, it is of great interest and importance to develop a novel imaging method for *in vivo* visualization of the spatial distribution of the SPIONs.

Magnetic approaches are the most promising techniques for *in vivo* imaging of SPIONs due to the unlimited penetration of magnetic signal. In the past years, magnetic resonance imaging (MRI) has been used to indirectly visualize the spatial distribution of the SPIONs via the effect of the SPIONs on proton relaxation [19]–[21]. However, the indirect imaging of the SPIONs with MRI suffers from several limitations [12], [22], [23], including the difficulty of SPION quantification and SPION-labeled cells discrimination in specific areas, as well as poor sensitivity. Magnetic particle imaging (MPI) is a new imaging modality, which is based on the direct measurement of the nonlinear magnetic response of SPIONs in magnetic fields [24], [25], thus allowing *in vivo* quantification and imaging of the SPIONs. Generally, an MPI scanner generates a gradient magnetic field – selection field (SF), saturating the SPIONs in a field-of-view (FOV) except in a field-free region (FFR), which is either a field-free point (FFP) or a field-free line (FFL) [24], [26]–[30]. In this case, only the SPIONs located in the FFR will response to external ac magnetic fields. By scanning FFR through the FOV, the magnetic responses of the SPIONs are measured for quantitative imaging. In principle, the sensitivity of MPI is about 2-3 orders better than that of MRI for the measurement of SPIONs due to higher susceptibility of the SPIONs [26].

To date, there have been various MPI approaches, differing in imaging methodology and scanner design. In 2005, the MPI technique was for the first time reported by Gleich and

Weizenecker, demonstrating the feasibility of MPI for the imaging of SPIONs [24]. The spectra of magnetic response of the SPIONs exposed in magnetic fields were measured for imaging with reconstruction, called $k$-space MPI. In 2009, Goodwill et al. proposed the methodology of $x$-space MPI, as well as the design of the x-space MPI scanner, which in principle allows the imaging of SPIONs without reconstruction [31]–[33]. In 2014, Vogel et al. developed an alternative approach of MPI, named traveling-wave MPI, without using permanent magnets [27]. In addition, single-sided MPI as developed to solve the problem of a size limitation for the specimens [34], [35] while FFL-based MPI approaches were also developed to improve the signal-to-noise ratio (SNR) [28], [36]–[39]. To date, MPI has been demonstrated to allow three-dimensional and real-time (46 frames per second) imaging of the SPIONs with a spatial resolution of about 1 mm [36]. These general MPI scanners are broadband approaches, which measure the full spectra of the SPIONs from tens to hundreds kilo-Hertz (even mega-Hertz). Other approaches of narrowband MPI via measuring a single harmonic have been reported for the imaging of the SPIONs with scanning of an FFP or FFL [40]–[46]. For instance, Pi et al. developed an MPI approach via the measurements of the SPION susceptibility [42]. In addition, other approaches using an array of magnetic sensors or mechanical scanning of a magnetic sensor or a phantom were also reported on for imaging SPIONs [47]–[49].

Different studies have also demonstrated the feasibility of MPI for multi-parametric imaging and biomedical applications. For instance, MPI has been used to visualize the spatial distribution of mobility [50]–[52], viscosity [53], [54], temperature [47], [55], [56] and binding behavior [57] in addition to the SPION concentration. For biomedical applications, MPI has shown its great promise in tumor imaging [58], magnetic hyperthermia [59], [60], targeted drug delivery [44], cell tracking [61], [62], perfusion imaging [63]–[65], real-time cardiac imaging [66], [67] and so on. Interestingly, Gräser et al built a human brain-sized MPI scanner [68], allowing the visualization of the spatial distribution of SPIONs in a human brain, which is capable of monitoring the cerebrovascular status.

The DF strength is one of the most important parameters in the design of an MPI scanner. In a general broad-band MPI scanner, the DF strength can be directly converted to the size of the imaging FOV with a given gradient magnetic field. To enlarge the FOV in a certain gradient, either the DF strength should be increased or an additional focus field should be applied to move the FFP/FFL. Generally, a higher DF strength allows stronger and more spectra of the SPIONs, thus improving the limit of detection (LOD) of the SPIONs. However, a higher DF strength need more power from a power amplifier, which will cause more harmonic distortion in the DF. The DF distortion can be directly fed to the receive signal from the SPIONs, which may worsen the LOD of the SPIONs. Especially when building a human-sized MPI scanner, the power and the total harmonic distortion (THD) will become a critical point. In addition, a higher DF strength may draw some safety issues, e.g. stimulation [69]. Furthermore, previous studies demonstrated that a lower DF strength may lead to a better spatial resolution [70], [71]. Thus, there is a trade-off regarding the DF strength. From this point of view, narrowband MPI is one of the promising approaches, balancing the DF strength to reach a good LOD and high spatial resolution.

In this paper, we propose an approach of single harmonic-based narrowband MPI to visualize the spatial distribution of the SPIONs. An MPI scanner is designed to generate a DF with frequency of 5 kHz for excitation and quasi-static fields to move the FFP through the FOV for imaging. The third harmonic at frequency of 15 kHz is measured for imaging with reconstruction. The DF strength-dependent point spread functions (PSFs) are measured and discussed. Experiments on different phantoms are performed to analyze the spatial resolution together with the PSFs.

## II. METHODS

### A. Concept

The concept of one-dimension (1D) MPI is depicted in Fig. 1. In a FOV, a gradient magnetic field $H_G$ generates an FFP, which can be moved by a scanning magnetic field $H_S$, as shown in Fig. 1a. Scanning the FFP position $x_{FFP}$ by $H_S$ (see Fig. 1b), the magnetic response of the SPIONs is modulated. Fig. 1c shows the amplitude of the third harmonic ($M_3$) of a spot SPION sample versus $x_{FFP}$ at frequency $3f_0$ in an excitation magnetic field with frequency $f_0$, which is considered as the PSF of the narrowband MPI. Fig. 1d shows a schematic of $M_3$ of three spot SPION samples versus $x_{FFP}$, which is a convolution of the PSF and the $M_3$ of each spot SPION sample. Note that the excitation magnetic field is set to be parallel to the sensitive axis of a detection coil. A deconvolution or reconstruction allows the measurement of the spatial distribution of $M_3$ generated by local SPIONs, which is proportional to the spatial distribution of the SPIONs. Therefore, scanning the FFP in multi-dimensions enables multi-dimensional imaging of the SPIONs.

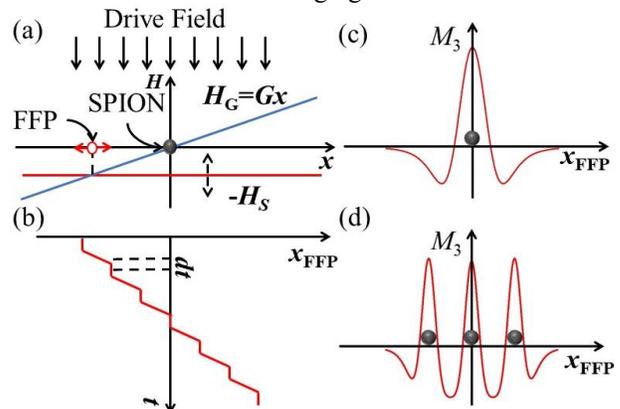

Fig. 1. Illustration of the single harmonic-based narrowband MPI. (a) shows the schematic of different magnetic fields for SPIONs excitation and FFP scanning. (b) shows FFP position vs. time. (c) shows the illustration of $M_3$ versus FFP position curve. (d) shows $M_3$ of three-spot SPION samples versus $x_{FFP}$ curve. Herein, the excitation magnetic field is set to be parallel to the sensitive axis of the detection coil.

### B. Scanner Design

According to the imaging concept, an MPI scanner was built to generate the desired magnetic fields and to measure the magnetic response of the SPIONs. Figs. 2a and 2b show a schematic and a photo of the MPI scanner, respectively.

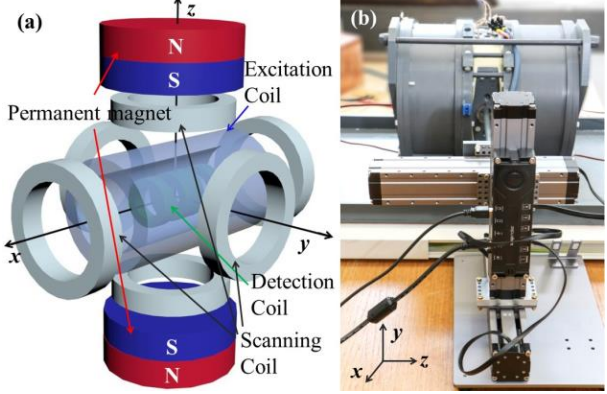

Fig. 2. Scanner overview. (a) shows the schematic of the MPI scanner. (b) shows the photo of the MPI scanner.

*1) Generation and scanning of the FFP*: The magnets of a former MPI scanner [30] were modified to generate the FFP and quasi-static scanning magnetic fields to move the FFP in the FOV, as shown in Fig. 2a. A pair of NdFeB permanent magnets were oppositely arranged to generate the FFP with a gradient of about 4.4 T/m in z-direction and 2.2 T/m in x- and y-directions. 3D Helmholtz/solenoid coils were used to generate 3D scanning magnetic fields $H_s$ for the FFP movement in the FOV. Fig. 2b shows the photo of the MPI scanner.

*2) Generation of the excitation magnetic field*: The excitation magnetic field $H_{ac}$ was generated by a solenoid coil in x-direction (see the transparent blue coil in Fig. 2a) at 5 kHz, driven by a power amplifier A 1110-16-A, purchased from Dr. Hubert (Bochum, Germany). Fig. 3 shows a schematic of the analog signal chain. A third-order LC bandpass filter was used to improve the THD of the excitation magnetic field. In addition, the impedance of the solenoid coil was matched by series and parallel capacitors, further improving the THD.

*3) Measurement of the SPION signal*: A gradiometric coil, including the detection and compensation coils, was used to detect the magnetic response of the SPIONs (see the green parts in Fig. 2a). The gradiometric coil was wound on a resin-printed bobbin with an inner diameter of 22 mm. In this paper, the 3$^{rd}$ harmonic ($M_3$) of the SPION signal at frequency 15 kHz was measured for imaging. To improve the SNR of the measurement signal and to suppress the feedthrough from the excitation magnetic field, a resonance circuit and a series capacitor were used to tune the resonance frequency of the gradiometric coil to 15 kHz, as shown in Fig. 3. Afterwards, a notch filter was used to further suppress the feedthrough signal. A low-noise preamplifier based on an AD8221 instrumentation amplifier was designed to amplify the signal, which was later on transferred to differential signals via an AD8066 and digitalized by an NI data acquisition card (NI PCIe-6374) with sampling rate of 1 MHz.

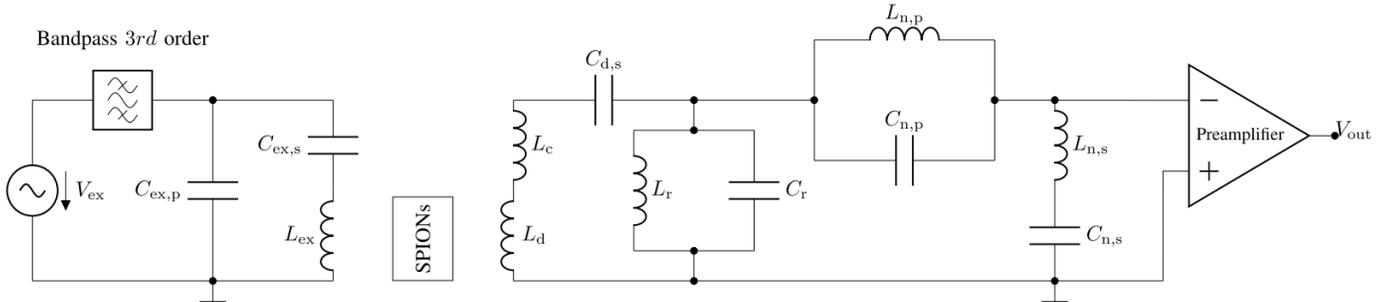

Fig. 3. Schematic of signal chain of the MPI scanner. The left side shows the transmit signal chain while the right side shows the detection signal chain.

### C. Scanner Test

The MPI scanner was tested with a spot SPION sample with 5 µL perimag stock solution, filled in a phantom with a dimension of 1.5 mm diameter in the xz plane and 3 mm length in y-direction. Fig. 4a shows the excitation and scanning magnetic fields versus time. Note that the excitation magnetic field is applied in x-direction while the FFP is scanned in z-direction with a quasi-static scanning magnetic field $H_s$ with a frequency of 1 Hz. It means that in 1 s the FFP is twice scanned through the FOV. A $H_s$ strength of 17.6 mT results in a FOV of 8 mm in z-direction. In this paper, the FFP was scanned in x-direction via mechanical movement by the positioning robot. Fig. 4b show the measured time-domain signals without (black curve) and with (red curve) the spot SPION sample. Note that the feedthrough from the excitation magnetic field was eliminated digitally. The inset in Fig. 4b shows the zoom-in signals from the MPI scanner. Fig. 4c shows the frequency spectra of the measured time-domain signal shown in Fig. 4b. Figs. 4b and 4c clearly show that without an SPION sample, the signal is dominated by the background noise and some residual feedthrough from the $H_{ac}$ Herein, the original feedthrough signal from the 3$^{rd}$ harmonic of the excitation magnetic field was already eliminated by a blank measurement. With an SPION sample, the 3$^{rd}$ harmonic was modulated by $H_s$, close to 15 kHz. In a time-interval of 0.0125 s, the FFP position was calculated from the scanning magnetic field while the $M_3$ of the spot SPION sample was measured with a digital lock-in amplifier. The calculated signal $M_3$ is depicted in Fig. 4d versus z-position, showing the 1D spatial distribution of the SPIONs.

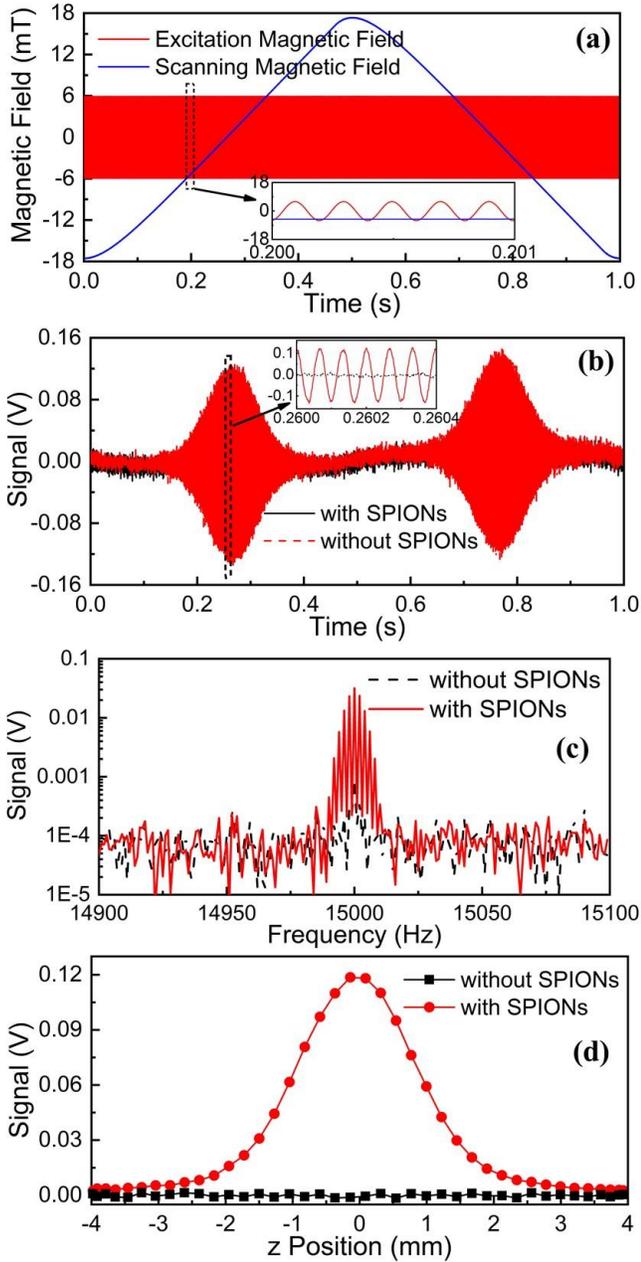

Fig. 4. (a) Excitation and scanning magnetic fields vs. time. Inset shows the zoom-in signals from 0.200 to 0.201 s. (b) Time-domain signals without (blank) and with (blue) spot SPION sample. Inset shows the zoom-in signals from 0.2600 to 0.2604 s. (c) Frequency spectra of the signals without (black) and with (blue) curves. (d) Measured 1D spatial distribution of the SPIONs, calculated from time-domain signals.

## III. RESULTS

### A. Experimental Description

In this paper, plain perimag SPIONs, purchased from micromod GmbH (Rostock, Germany), are used as experimental sample. Perimag SPIONs with a hydrodynamic diameter of about 130 nm are multi-core particles, consisting of several iron-oxide cores in a dextran matrix. The iron concentration of the stock sample is 8.5 mg/mL. In this paper, 2D images of the spatial distribution of SPIONs are measured in the xz-plane with the MPI scanner. The measurement time of one line in z-direction amounts to 1 s (1 s data acquisition).

Each pixel of all the measured images has a dimension of 0.2 mm × 0.2 mm. All the measured 2D images of the phantom were reconstructed from raw data with a regularized Kaczmarz method based on a measured system matrix [72].

### B. PSF measurement

To evaluate the spatial resolution, the same spot sample for the MPI scanner test was used to measure the PSFs at different $H_{ac}$, as shown in Fig. 5. The FOV for the PSF measurement is 12 mm × 8 mm in x- and z-direction. It clearly indicates that with increasing excitation magnetic field the measured signal area gets larger, representing a worsened spatial resolution.

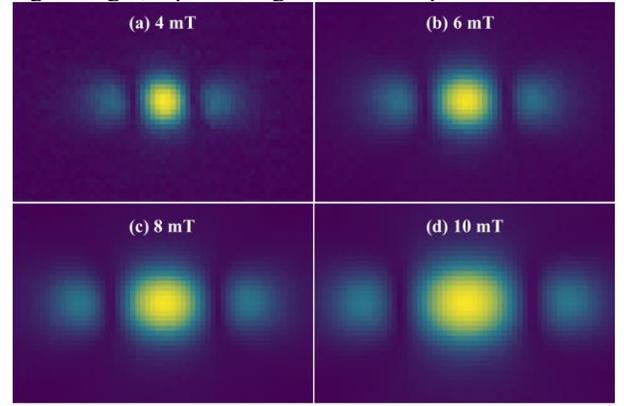

Fig. 5. Measured 2D images of PSFs at different excitation magnetic fields of 4 mT (a), 6 mT (b), 8 mT (c) and 10 mT (d). The image intensity is normalized to 1.

To compare the spatial resolutions at different $H_{ac}$, the 1D PSFs in x- and z-direction are presented in Figs. 6a and 6b. They clearly show that the PSF gets broader with increasing $H_{ac}$ both in x- and z-direction. To quantitively analyze the spatial resolution, the full width at half maximum (FWHM) of the PSF was calculated and presented in Fig. 5c. Note that a half of the difference between the maximum and minimum values is used to calculate the FWHM. With increasing the amplitude from 4 to 10 mT, the FWHM in x-direction increases from 2 to 4.4 mm while the FWHM in z-direction rises from 1.8 to about 3.0 mm.

### C. Phantom Images

Two-line phantoms in x- and z-direction with different distances were used for experiments to evaluate the spatial resolution of the narrowband MPI. Each line has 1 mm width, 4 mm length in the xz plane and 3 mm depth in y-direction. Each line was filled with 12 μL perimag SPIONs. The distance between the two lines (center-to-center) was varied from 1.5 to 3 mm (edge-to-edge distance from 0.5 to 2 mm). Note that in this paper, the line-to-line distance is from center to center if not specified otherwise. The measured FOV amounts to 14 mm × 8 mm in x- and z-direction. In addition, different $H_{ac}$ from 4 to 10 mT were applied to investigate the effect of the $H_{ac}$ strength on spatial resolution whereas the SPION concentration was varied to investigate the effect of the measurement SNR on spatial resolution.

Figure 7 shows the reconstructed 2D images of the two-line phantoms filled with 8.5 mg/mL stock suspension. It indicates that the two lines with center-to-center distances of 3.0 mm and 2.5 mm can be clearly distinguished at all different $H_{ac}$. For a line-to-line distance not larger than 2.0 mm, the images got blurred, especially at a stronger $H_{ac}$ and in x-direction.

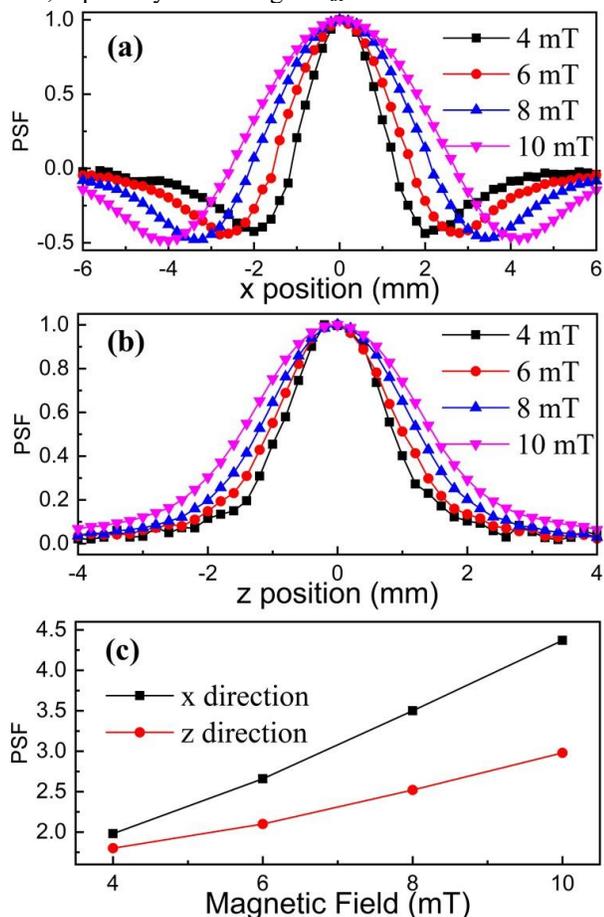

Fig. 6. (a) and (b) show the normalized 1D PSF in x- and z-direction. (c) shows the dependence of the FWHMs on the amplitude of the excitation magnetic field in x- and z-direction. Symbols are experimental data whereas solid lines are guides to the eye.

Fig. 8 shows the line scans in the center of the reconstructed 2D images shown in Fig. 7. The upper (lower) row shows the normalized intensity versus position (intensity-position curve) in x- (z-) direction. Each column shows the intensity-position curve at different $H_{ac}$. For two lines with distances of 3.0 and 2.5 mm, the reconstructed 2D images in Fig. 7 and the intensity-position curves (black and red curves) in Fig. 8 show two lines and peaks, respectively. It means that the spatial resolution is better than 3.0 mm both in x- and z-direction. For two lines with a distance of 2.0 mm, in x-direction, the reconstructed 2D images in Fig. 7 and the intensity-position curves in Fig. 8 at 4, 6 and 8 mT show two lines and two peaks. However, at 10 mT, the reconstructed image does not clearly show two lines whereas the intensity-position curve (blue curve in Fig. 8(d)) shows two peaks. It may indicate that the spatial resolution at 10 mT in x-direction might be 2.0 mm. In z-direction, two lines and two peaks can be seen from the reconstructed 2D images in Fig. 7 and intensity-position curves (blue curves) in the upper row in Fig. 8, indicating a spatial resolution better than 2.0 mm in z-direction. For two lines with a distance of 1.5 mm, at 4 mT the reconstructed 2D images and intensity-position curve show two lines and two peaks, indicating a spatial resolution of 1.5 mm both in x- and z-direction. At 6 mT, in the z-direction, two lines and two peaks can be seen from the reconstructed 2D images and the intensity-position curves (magenta curves) in Figs. 8(b) and 8(f). However, in x-direction, the reconstructed 2D image and intensity-position curve cannot resolve two lines or two peaks, respectively. It means that the spatial resolution is worse than 1.5 mm in x-direction. At 8 and 10 mT, all the reconstructed 2D images and intensity-position curves cannot resolve the two lines.

In addition, the same phantom experiments were also performed with different SPION concentrations ranging from 8.5 to 1.1 mg/mL (experimental results not presented here). They indicate that the measured images contain an increasing number of shadows/artifacts with decreasing iron concentration due to a worsened SNR. Moreover, it also becomes more difficult to distinguish the two lines from the phantom, especially with a small distance, e.g. of 1.5 mm at 4 mT. In general, the phantom experiments demonstrated that the spatial resolution in z-direction is better than that in x-direction. Moreover, the spatial resolution in both x- and z-direction gets worse with increasing the excitation magnetic fields.

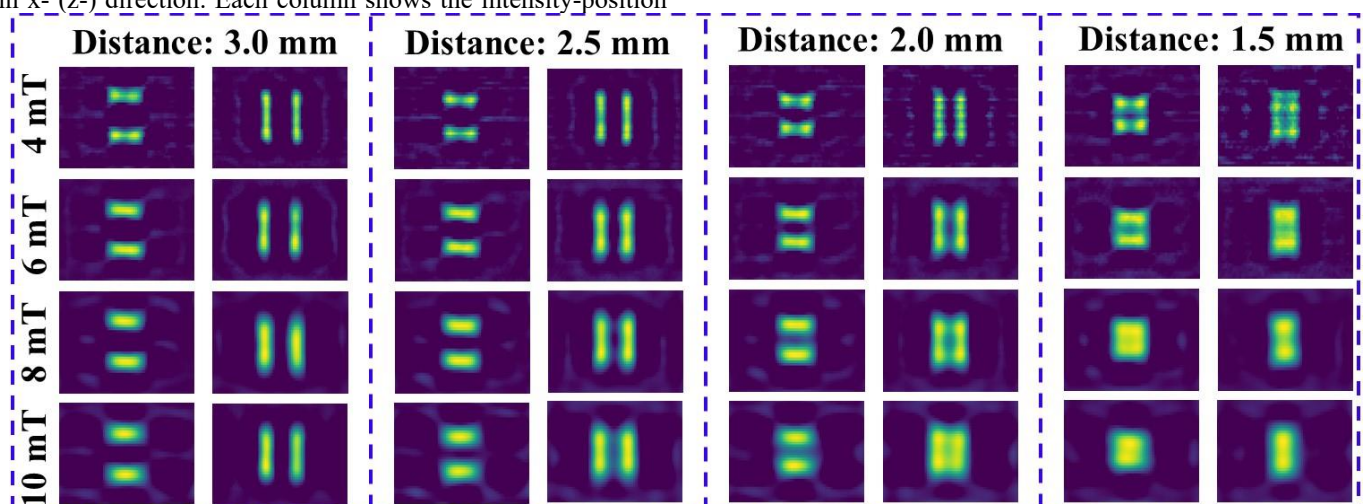

Fig. 7. Measured 2D images of two-line phantoms at different magnetic fields with different distances. The iron concentration is 8.5 mg/mL. The measured FOV is 14 mm × 8 mm in x- and z-direction.

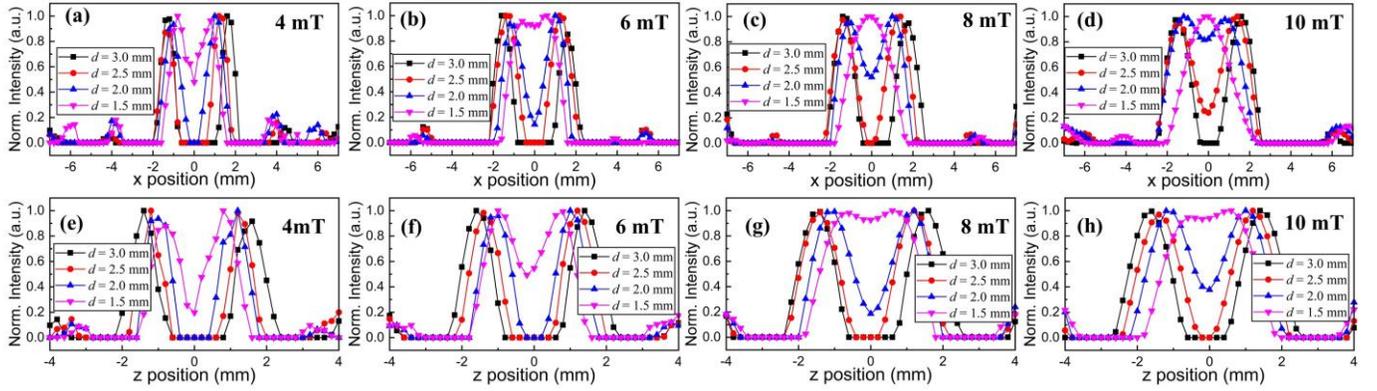

Fig. 8. Experimental results of normalized image intensity vs position curves of the phantoms with 8.5 mg/mL. Symbols are experimental data whereas solid lines are guides to the eyes.

### D. Modulation Transfer Function

The modulation transfer function (MTF) is used as another metric to quantitatively evaluate the spatial resolution of the narrowband MPI. The MTF is defined as $MTF(f_{spatial}) = (c_{max} - c_{min}) / (c_{max} + c_{min})$, where $c_{max}$ ($c_{min}$) is the maximum (minimum) value of the 1D curve at a given spatial frequency $f_{spatial} = 1/d$ with the center-to-center distance $d$ between two lines [73]–[75]. Fig. 9 shows the calculated MTF vs. spatial frequency $f_{spatial}$ at different $H_{ac}$ for different-concentration SPION samples, calculated from reconstructed images. Note that Fig. 9 only presents the MTFs for three different concentrations. The spatial resolution $R$ is defined as the value at which $MTF(1/R)$ equals 0.5. As can be seen in Fig. 9, with increasing the excitation magnetic field, the $1/R$ at which MTF = 0.5 decreases, corresponding to a worsening of the spatial resolution $R$ with increasing field amplitude.

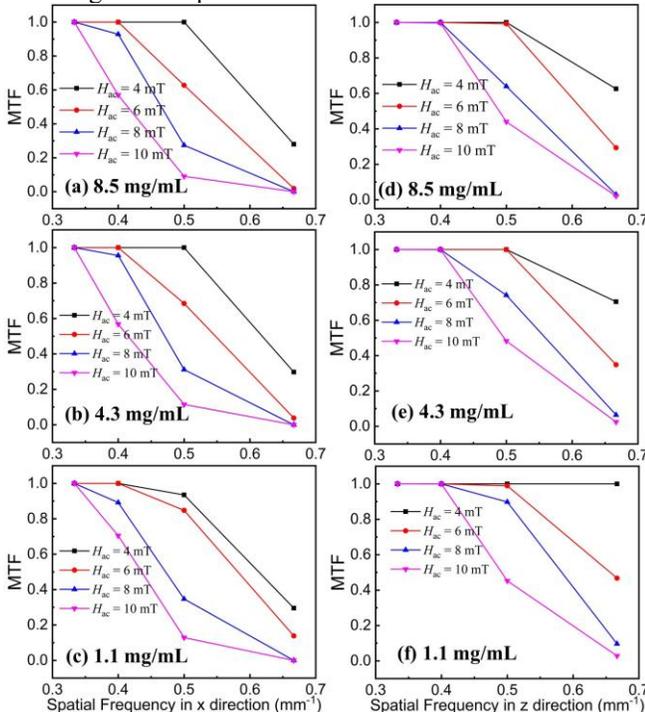

Fig. 9. Experimental results of modulation transfer function in x- and z-directions at different excitation magnetic fields for three different-concentration SPION samples.

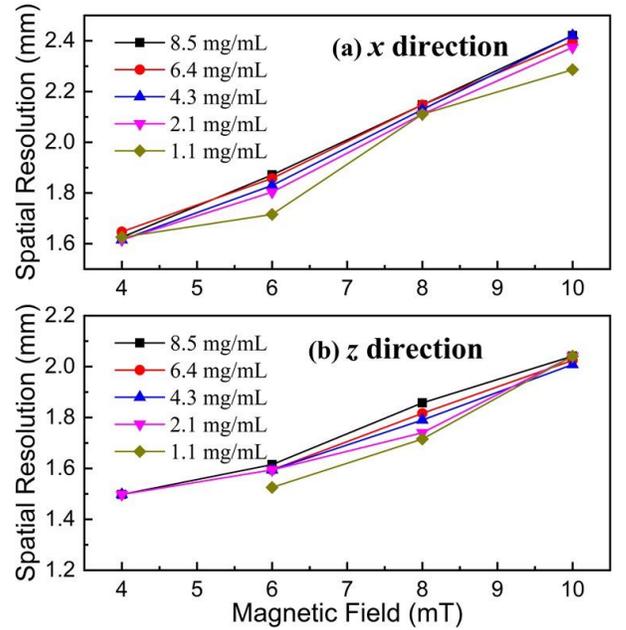

Fig. 10. Spatial resolution at different magnetic fields in x- (a) and z- (b) directions, calculated from the MTF. Symbols are experimental results whereas solid lines are guide to the eye.

### E. Limit of Detection

The limit of detection (LOD) is one of the most important measures in an imaging modality. To quantitatively assess the LOD of the custom-built narrowband MPI scanner, a point phantom with 2 mm diameter in the xz plane and 3 mm length in y direction was filled with 9.4 μL perimag SPIONs with different concentrations ranging from 8.5 mg/mL to 0.053 mg/mL (iron weight ranging from 80 μg to 500 ng). The phantom was imaged at different $H_{ac}$ with a FOV of 4 mm × 4 mm in the xz plane.

Figure 11 shows the reconstructed 2D images of the point sample. It shows that the spot sample is discernable down to 425 μg/mL at $H_{ac}$ = 4 mT. At $H_{ac}$ = 6 mT and 8 mT, the measured images exhibit a clear spot down to 106 μg/mL, but show strong artifacts for 53 μg/mL. At $H_{ac}$ = 10 mT the spot sample is detectable down to 53 μg/mL. The comparison of the reconstructed images for 53 μg/mL and 0 μg/mL at 10 mT indicates a better LOD than 53 μg/mL. Therefore, according to the measured and reconstructed images, the LOD of the current MPI scanner is about 53 μg/mL, corresponding to 500 ng in terms of iron weight.

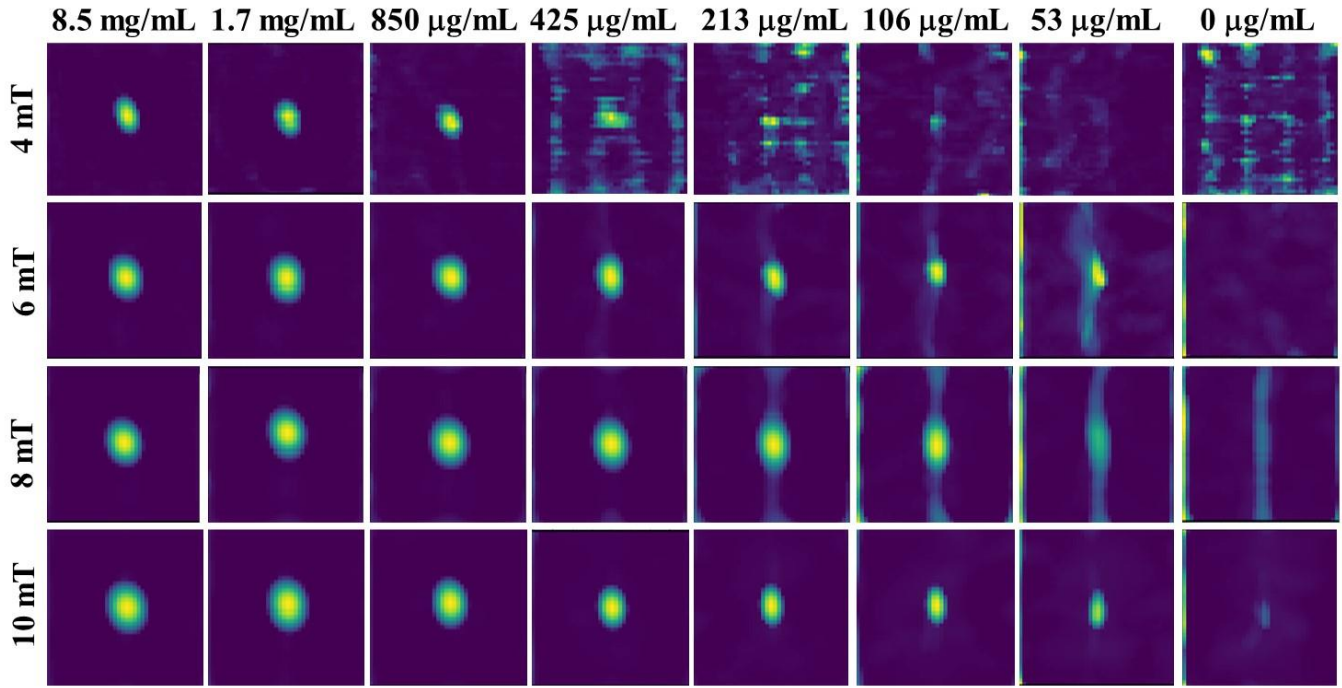

Fig. 11. Reconstructed images of a point phantom with different-concentration SPION samples at different excitation magnetic fields.

### F. Logo Image

To further demonstrate the performance of the narrowband MPI, a "complex" phantom – our institute logo "EMG" – was filled with 8.5 mg/mL perimag SPIONs for experiments. Fig. 12 shows the photo of the emg logo phantom and the reconstructed image. The imaging FOV is 34 mm × 10 mm in x- and z-direction. Considering the spatial resolutions and the SNR at different $H_{ac}$, the image was measured at $H_{ac}$ = 6 mT. The reconstructed image, as shown in Fig. 12b, can clearly resolve the designed logo. At the upper and downer edges in z-direction, there are some blurring in addition to the artifacts, which may come from some SPIONs at the edge (see Fig. 12a) during preparation. In addition, some signal dropout happened in the letter "e" (see the white dash arrow). Nevertheless, the logo image demonstrates the feasibility of the narrowband MPI for SPION imaging.

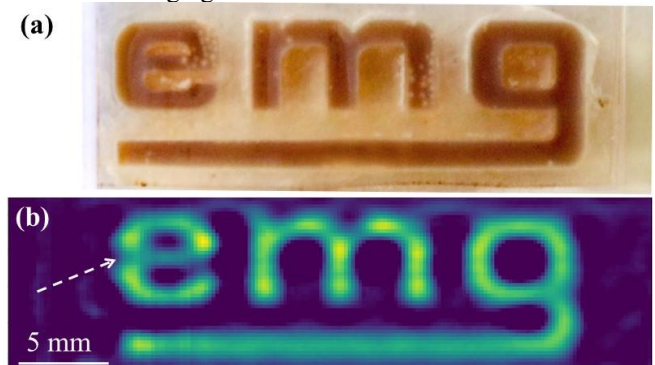

Fig. 12. (a) Photo of the emg-logo phantom image filled with SPIONs. (b) Reconstructed images of the emg logo.

## IV. Discussion

The spatial resolution and LOD are the most important measures for an imaging modality, thus for MPI as well. To date, the spatial resolution of MPI is in the range of about 1 mm or submillimeter, which depends on the gradient of the selection magnetic field and the shape of the SPION magnetization curve [31]. Note that one should keep in mind that the spatial resolution is defined as the distinguishable minimum gap between two lines (edge-to-edge distance) or the center-to-center distance. In terms of a rigorous definition, the center-to-center distance should be used to evaluate the spatial resolution whereas in practice, the edge-to-edge distance (two-line gap) is easy to quantify. For instance, the minimum distinguishable gap between two lines in this paper is 0.5 mm at $H_{ac}$ = 4 mT, which corresponds to a spatial resolution of 1.5 mm considering the line width. Note that the FWHM of the PSF only gives an estimation of the spatial distribution. A reconstruction significantly improves the spatial resolution.

The spatial resolution is dependent on not only the gradient magnetic field, but also the direction and strength of the excitation magnetic field, as well as the SPION dynamics. In this paper, the gradient in z-direction is two times higher than that in x-direction, which in principle predicts a two-fold better spatial resolution in z-direction. However, the experimental results show that the spatial resolutions in x- and z-direction were similar, at least not two-fold different. It may be caused by the relaxation behavior and the direction of the excitation magnetic field, which qualitatively fits with other studies [42], [76]. With an excitation magnetic field only in x-direction, the spatial resolution was better in x-direction than in the other directions if the gradients were the same in all directions. In this paper, we found that with increasing the excitation magnetic field, the spatial resolution gets worse, which fits quite well with previous studies [70], [71]. The underlying physics are rather complicated. In principle, the relaxation behavior of the SPIONs blurs an MPI image [76]. Thus, increasing the excitation magnetic field or drive field strength decreases the relaxation time of the SPIONs and consequently results in a faster rotation [77], [78], which in principle compensates the blurring caused by the relaxation behavior. However, increasing the excitation magnetic field or drive field strength, the PSF gets broader, thus worsening the spatial resolution. It means that there may be an optimum strength of the excitation magnetic field or drive field. This optimum strength should also depend on the excitation frequency, as well as the SPION properties. In broadband MPI approaches, e.g. k-space and x-space MPI, the full spectra of the magnetic response from the SPIONs should be measured for imaging. Therein, higher harmonics are very important for obtaining sharp edges (a better spatial resolution) of an image, but also may worsen the SNR due to a smaller amplitude of the higher harmonics. On the contrary, lower harmonics allows better SNRs but a worse spatial resolution. A stronger ac magnetic field in principle improves the SNR. However, one should keep in mind that it not just worsens the spatial resolution, but also significantly increases the required power to drive the coils. Thus, it may cause more distortions in the excitation magnetic field and the detection system due to direct feedthrough, thus worsening the LOD. A weaker ac magnetic field will significantly decrease the FOV.

In the presented narrowband MPI, only the 3rd harmonic of the SPION signal was measured for imaging. It may have a comparable good balance between the LOD and spatial resolution in a given excitation magnetic field. In addition, the required bandwidth of the detection system is in the order of a few tens of Hz (see Fig. 4c). A resonance circuit was employed to improve the sensitivity of the detection system, thus allowing a quite good LOD even at low magnetic fields, e.g. 6 mT, and low excitation frequency, e.g. 5 kHz. In principle, with increasing the excitation frequency, the SNR can be further improved due to Faraday induction law, as well as a better Q-factor of the detection coil. In this case, it will be possible to have a better trade-off between the spatial resolution and the LOD. The efforts on hardware implementation, e.g. filter design in the transmit side, when increasing the excitation frequency in narrowband MPI is much less than those in broadband MPI. In addition, the concerns in magnetostimulation in narrowband MPI are also less crucial than in broadband MPI due to a lower excitation magnetic field.

In the broadband k-space and x-space MPI approaches, the FOV significantly depends on the strength of the drive magnetic field with a given gradient magnetic field. The generation of the drive magnetic fields with high frequencies, e.g. 25 kHz, and high amplitudes, e.g. 20 mT, is quite challenging by taking into account the THD. In the presented narrowband MPI, the FOV mainly depends on the quasi-static magnetic field for FFP scanning, but independent of the excitation magnetic field at 5 kHz. The efforts to generate the quasi-static magnetic fields are significantly reduced in terms of hardware implementation, e.g. the required filters to improve the THD in the transmit side. Some of these advantages are based on the sacrifice of the temporal resolution. For instance, the quasi-static magnetic fields for the FFP scanning results in a worse temporal resolution compared to standard k-space MPI. Considering all the factors involved in MPI, e.g. spatial resolution, sensitivity and safety, we believe that single-harmonic-based narrowband MPI may be a quite promising approach.

## V. Conclusion

In this paper, we systematically investigated a single-harmonic based narrowband MPI. The concept of the narrowband MPI was introduced while a narrowband MPI scanner was built to visualize the spatial distributions of SPIONs. Different phantom experiments were performed to demonstrate the feasibility of the narrowband MPI for SPION concentration imaging. In addition, FWHM and MTF were measured and used as different metrics to evaluate the spatial resolution, as well as its dependence on the strength of excitation magnetic field. The LOD of the custom-built MPI scanner was determined based on images of phantoms with SPIONs with different concentrations. We envisage that the present methodology is an alternative MPI approach, showing promising performance in terms of spatial resolution and LOD. We believe that our study is of great interest and importance to biomedicines, including molecular imaging, cell tracking and targeted drug delivery.

## Acknowledgement

Financial support from German Research Foundation (DFG) under Grant ZH 782/1-1 is gratefully acknowledged.